\documentclass[superscriptaddress,groupedaddress,nofootnoteinbib,11pt]{article}
\pdfoutput=1 
\usepackage{jheppub}

\usepackage{CJKutf8} 
\usepackage{mathtools,slashed,mathrsfs}
\usepackage[caption=false]{subfig}
\usepackage{dcolumn}
\usepackage{multirow}
\usepackage{tabularx}
\usepackage{booktabs}
\usepackage{bm}
\usepackage{comment}
\usepackage{setspace}
\usepackage[dvipsnames]{xcolor}
\usepackage[normalem]{ulem} 
\usepackage{enumerate}
\usepackage{siunitx}
\usepackage{graphicx}
\usepackage{slashed}

\definecolor{mypurple}{RGB}{164,64,214}

\newcommand\nn{\nonumber}
\newcommand\eea{\end{eqnarray}}
\newcommand\bea{\begin{eqnarray}}

\newcommand{\G}{\gamma_{N}}

\def\l{\left(}
\def\r{\right)}
\def\la{\langle}
\def\ra{\rangle}

\newcommand{\newc}{\newcommand}
\def\be{\begin{equation}}
\def\ee{\end{equation}}
\newc{\gsim}{\lower.7ex\hbox{$\;\stackrel{\textstyle>}{\sim}\;$}}
\newc{\lsim}{\lower.7ex\hbox{$\;\stackrel{\textstyle<}{\sim}\;$}}

\def\luv{\Lambda_{{\scriptscriptstyle {UV}}}}

\begin{document}

\newcommand{\ric}[1]{{\color{blue}#1}}

\title{
Softening the UV without New Particles
}
%

\author[a]{Anson Hook,}
\author[b]{and Riccardo Rattazzi}

\affiliation[a]{Maryland Center for Fundamental Physics, University of Maryland, College Park, MD 20742}
\affiliation[b]{Institute of Physics, Theoretical Particle Physics Laboratory (LPTP), \\ 
\'Ecole Polytechnique F\'ed\'erale de Lausanne (EPFL), 
CH-1015 Lausanne, Switzerland}
\emailAdd{hook@umd.edu}
\emailAdd{riccardo.rattazzi@epfl.ch}

\abstract{

We explore an odd class of QFTs where a hierarchy problem is resolved with new dynamics as opposed to new particles. The essential element of our construction is a $U(1)$  pseudo-NG boson with symmetry breaking interactions all characterized by a large number $N$ of units of the fundamental charge. In the resulting effective theory, quantum corrections, like those to the effective potential and mass, which are normally power divergent and saturated at the UV cut-off, are instead saturated at a much lower scale.
This critical scale, which does not involve  any new particle, corresponds to the onset of unsuppressed multiparticle production in scattering processes. Remarkably this all happens within the tractable domain of weak coupling. Terms involving arbitrarily high powers of the Goldstone field must  however be taken into account. In particular, a truncation to the renormalizable part of the effective Lagrangian would completely miss the physics.


}

\maketitle

\section{Introduction}

The existence of large separations of length scales is a basic fact of Physics.  Indeed, our ability to describe phenomena in terms of a finite number of parameters is a direct consequence of that fact. More precisely what controls predictivity  is the separation  between the fundamental scale of the dynamics and the macroscopic scale of the phenomenon being described. The multipole expansion in classical electrodynamics represents the simplest incarnation  of the concept.
Modern effective field theories (EFTs), with their action organized as an expansion in a series of operators of increasing dimension, are just a more sophisticated one.  The added difficulty in the latter case stems from ultraviolet (UV) divergences. These can be  technically dealt with via the renormalization procedure, in a manner that is independent of the specific nature of the microphysics. In particular the scales that  regulate the UV divergences in a chosen scheme do not need to be physical. In reality, however, we expect these divergences to be regulated or at least modified at the physical scale where the EFT gives way to a more fundamental description. Of particular relevance are, in this perspective, divergences that grow with a power of the UV cut-off. That is because these are normally associated with
corrections to the physical masses and  thus control the very existence of the separation of scales that makes the EFT description possible. 

In all known examples where the mass of a scalar  is UV completed into a theory where it is calculable, the fate of the power divergence of the EFT is invariably the same: 
the order of magnitude of the physical effects are correctly captured by just cutting-off the UV divergences of the EFT at momenta around the physical mass of the particles of the UV completion. There are many real world and theoretical examples.  Low energy QCD offers two real world examples with the $K - \bar K$ and  $\pi^+-\pi^0$ mass differences. In the $K - \bar K$ case, the relevant EFT involves the charged-current Fermi
interaction of $u,d,s$-quarks supplemented with Cabibbo mixing and is UV completed by the addition of the charm quark $c$. Neglecting $O(1)$ factors, one has the pattern
\be
\Delta m_K^2\sim \left [\frac{G_F^2 f_K^2m_K^2}{16\pi^2}\right] \luv^2 \,\rightarrow\, \left [\frac{G_F^2 f_K^2m_K^2}{16\pi^2}\right] m_c^2 
\label{KK}
\ee
where the first result represents the estimate in terms of the EFT cut-off $\luv$, while the second is the correct calculation in the UV completed theory~\cite{Glashow:1970gm,Gaillard:1974hs}. Of course this result crucially depends on the absence of the power divergence in the UV completed theory. In the case of the $\pi^+-\pi^0$ mass difference, the EFT is the chiral Lagrangian for pions supplemented with electromagnetic interactions. The UV completion includes the heavier hadrons, whose mass can collectively be identified with that of the $\rho$ meson $m_\rho$. Again up to $O(1)$ factors and with the same notation as before one has
\be
m_{\pi^+}^2-m_{\pi^0}^2\sim \left [ \frac{e^2}{16\pi^2}\right ] \luv^2  \,\rightarrow\, \left [ \frac{e^2}{16\pi^2}\right ] m_\rho^2\,.
\label{pipi}
\ee

Models with a calculable Higgs mass, such as supersymmetric~\cite{Martin:1997ns},  or composite~\cite{Panico:2015jxa} models,  all offer   theoretical, but conceptually robust, incarnations of the same situation. For instance the contribution of the top quark sector to the Higgs mass parameter invariably follows the pattern
\be
\delta m_H^2 \sim \left [\frac{3y_t^2}{4\pi^2}\right ] \luv^2\,\rightarrow\, \left [\frac{3y_t^2}{4\pi^2}\right ] m_T^2
\label{deltamh}
\ee
where $m_T$ represents the physical mass of the top partners, which are  bosons in the case of supersymmetry and fermions in the case of composite Higgs \footnote{The top partners normally carry the same color quantum number as the top. Twin Higgs models offer, at the price of additional complication, a twist where the lightest  top partners do not carry color~\cite{Chacko:2005pe}.  But even in that case eq.~(\ref{deltamh}) holds true.}.

In the  simplest cases,  like in eqs.~(\ref{KK},\ref{pipi}), power divergences imply a remarkable relation, valid up to an $O(1)$ factor,  between parameters that are measurable at low energy within the  EFT and the scale where the EFT breaks down.  In more general situations the $O(1)$ factor represents just an upper bound, as one can engineer cancellations between different UV contributions.   These cancellations can be seen explicitly in theories such as supersymmetry and composite Higgs. For instance in the case of supersymmetry, the cancellation in the Higgs mass can occur between the contribution due to sparticles and the tree level contribution from the $\mu$ term.  Significant cancellations, however, appear  non-generic, and thus un-natural. The reason for that view is that, while the structure of the individual contributions is  robustly based on symmetry and selection rules, their cooperative cancellation  is not. The relevant symmetry constraints are implicit in the structure of eqs.~(\ref{KK},\ref{pipi},\ref{deltamh}). These all involve two factors: the first, within  brackets, corresponds to the square of a dimensionless coupling constant, while the second is just the square of a physical mass \footnote{This statement is manifestly correct for eqs.~(\ref{pipi},\ref{deltamh}) while for eq.~(\ref{KK}) 
 notice that the prefactor can be rewritten as $(g^2/16\pi^2)(m_K/m_W)^2(f_K/v_F)^2$, which more precisely corresponds to a squared coupling times ratios of masses and decay constants.}. The appearance of each factor is dictated by a separate set of selection rules. The second factor, the squared mass, is just dictated by the dilation symmetry selection rule, a.k.a. dimensional analysis. The  first factor is dictated by the selection rules of the group of  higher spin symmetries of free field theory \footnote{In the Appendix we offer a more detailed discussion of this fact (see also \cite{RR2}). }. Under this symmetry, all couplings (i.e. all the coefficients of higher than quadratic terms in the action) can be viewed as spurions with non-trivial transformation properties. It is the selection rules associated with these transformations that dictate the presence of the squared coupling factor. This symmetry explains why, in all UV completions including the top-Higgs coupling,  there always appears a correction of the form shown in eq.~(\ref{deltamh}). The point is 
 that, whenever the Yukawa coupling $y_t$ and the mass scale $m_T$ exist in the theory, a correction of the form
 in  eq.~(\ref{deltamh}) is allowed, given  it matches  the quantum numbers of the Higgs mass term under (higher spin symmetry) $\times$ (dilation). 

Given the important role hierarchy problems play in particle physics, it is important to  study  any possible exception to the naivest interpretation of naturalness.  
Along these veins, in this paper we study precisely one such exception~\footnote{There have been many attempts to circumvent the arguments of naturalness, see Refs.~\cite{Weinberg:1987dv,Graham:2015cka,Arkani-Hamed:2016rle,Hook:2018jle,Geller:2018xvz,Cheung:2018xnu,Giudice:2019iwl,Arkani-Hamed:2020yna,Giudice:2021viw,TitoDAgnolo:2021nhd} for a representative sample.  Similar to this paper, there have also been recent attempts at violating the expectations of naturalness~\cite{Creminelli:2005ej,Hook:2019mrd,Arkani-Hamed:2021xlp}.}. 
We will present and study a toy model where power divergences in the EFT are rendered finite rather surprisingly at a scale that is parametrically below  where new resonances appears. 
That is as if the role of $m_T$ in eq.~(\ref{deltamh})  was played in reality by a scale $\lsim 500$ GeV at which  no new states exist, compatible with naturalness and compatible with the lack of direct evidence for top partners below $\sim 1-2$ TeV. Surprisingly, the symmetry rendering all of this possible is a simple discrete shift symmetry.  The leading order corrections to the mass coming from a shift symmetric Yukawa or from a shift symmetric scalar potential are finite.

While violating the naivest interpretation of naturalness, our example does not violate a more refined definition of naturalness.  Before the scale $\luv$ new physics does occur, just not in the form of new particles.  Instead, final states with multiple particles become important.  Nonetheless the systems remains weakly coupled and tractable: the dominant final states contain a large but finite number of quanta and the cross section is perturbatively small.
The virtual counterpart of this on-shell phenomenon is responsible for the finiteness of the relevant class of loop integrals.

Another interesting feature in our model is that the correction to the scalar mass is algebraically related to the mass of the Yukawa coupled fermion.  In our simple example, we find that $\delta m_\phi = 2 m_\psi$.  While we are unsure of what exactly are the full implications of such a relationship, it is amusing to note that, to within a few percent, $m_H = m_t / \sqrt{2}$.

Exceptions almost invariably come at a price.  In our case the price is an extra parameter that allows us to lower the loop cut-off below the masses of the new states. As made evident by considering the UV completion of the EFT,  this parameter is essentially the large number of legs, or the large charge, and thus the large dimensionality of the involved operators~\footnote{The fact that the operators involve fields to high powers is why multiple final states become important.  The theory must interpolate from the deep IR where there is a simple Yukawa coupling involving a single particle, to the far UV where this Yukawa coupling is a higher dimensional operator and involves many particles.}. 
The large dimensionality of the operators that are involved makes it difficult to extend our mechanism to larger couplings.
Because of this, a translation of the mechanism of our toy model into a concrete and natural UV completion of the Higgs does not seem immediate.
Perhaps, and that is our hope, others may succeed in putting the mechanism to good use by working along equally unusual pathways.
Given that most alternative explanations for the Higgs mass have been experimentally cornered, perhaps nature is indicating a theory of this kind is present.

This paper is organized as follows.  In Section~\ref{Sec: EFT} we discuss our theory and show that a shift symmetry is all that is needed to render some would be divergent diagrams finite.  In Section~\ref{Sec: UV} we present a UV completion and show that its calculations agree with those in Section~\ref{Sec: EFT}.  Finally we conclude in Section~\ref{Sec: conclusion}.

\section{IR EFT} \label{Sec: EFT}

In this section we study the IR EFT of interest.  Our starting point is a Pseudo-Nambu Goldstone boson with a discrete ${\mathbb Z}_2\rtimes {\mathbb Z}$ symmetry 
\be
{\mathbb Z}_2:\,\phi\rightarrow -\phi\,,\qquad\qquad {\mathbb Z}: \phi \rightarrow \phi + 2\pi k f \quad k\in  \mathbb{Z}\,.
\ee
 The  Lagrangian we will consider is
\bea
\label{eq: IR EFT}
\mathcal{L} = \frac{1}{2} \l \partial \phi \r^2 + i \overline \psi\slashed\partial \psi+\epsilon^4 \cos \l \frac{\phi}{f} \r + \sqrt{2} y f \sin \l \frac{\phi}{2f} \r \overline \psi \psi.
\eea
Invariance of the Yukawa coupling  under ${\mathbb Z}_2\rtimes {\mathbb Z}$  dictates  $\psi$ transform in such a way that ${\mathbb Z}_2: \overline \psi \psi \rightarrow (- 1)\overline \psi \psi$,   ${\mathbb Z}: \overline \psi \psi \rightarrow (- 1)^k\overline \psi \psi$.  The above Lagrangian involves the {\it lower harmonics} in $\phi$ that are compatible with ${\mathbb Z}_2\rtimes {\mathbb Z}$. Higher harmonics, i.e. terms involving higher powers of $\cos \l \frac{\phi}{f} \r$, are generated at loop level but are correspondingly suppressed by higher powers of the couplings $\epsilon^4$ and $y$, which we treat as small. While this specific structure isn't necessary, it is convenient.

The  higher harmonics generated by quantum corrections are indeed the main target of our discussion.
At first glance, many loop integrals are expected to be divergent, e.g. a loop of fermions giving a mass to $\phi$.  However, a more detailed calculation demonstrates that these loops are instead regulated by the scale $4 \pi f$.  While the request of perturbative unitarity\footnote{We stick to the traditional nomenclature, even though it is inaccurate, because unitarity never is at stake:  these bounds simply require  the theory to be weakly coupled, so that it makes sense to write a Lagrangian.}   typically implies the  UV cutoff of theories of Goldstone bosons to be below the scale $4 \pi f$, it seems clear that the above theory can have a UV cutoff above $4 \pi f$.  As $\epsilon$ and $y$ go to zero, the theory becomes free  and the UV cutoff, as dictated by unitarity, goes to infinity.

In this section, we first calculate the unitarity bounds associated with the Lagrangian in eq.~(\ref{eq: IR EFT}) before calculating a few loop diagrams and showing that they are regulated at the  scale $4 \pi f$.  

\subsection{Unitarity Bounds}

We first calculate the unitarity bound associated with the potential, $ \epsilon^4 \cos \l \frac{\phi}{f} \r$.
We will follow the approach of Ref.~\cite{Chang:2019vez} with results in theories similar to ours being found in Refs.~\cite{Falkowski:2019tft,Craig:2019zkf,Ekhterachian:2021rkx}.  
We consider an initial state with $n$ Goldstone bosons scattering to a final state with $n$ Goldstone bosons (states with different number of initial and final states give weaker unitarity bounds).
The appropriate dimensionless matrix element is
\bea
\mathcal{M}_{n \rightarrow n} = \frac{1}{8 \pi} \frac{\epsilon^4}{f^4} \left ( \frac{\sqrt{s}}{4 \pi f} \right )^{2n - 4} \frac{1}{n! (n-1)! (n-2)!} \approx \frac{\epsilon^4}{n^6 f^4}  \left ( \frac{\sqrt{s}}{4 \pi f n^{3/2}} \right )^{2n - 4} .
\label{Mn}
\eea
In the last line, we have taken the large $n$ limit.  This matrix element is maximized when scattering $n_\text{max}$ particles,
\bea
n_\text{max} = \frac{1}{e} \left ( \frac{\sqrt{s}}{4 \pi f}  \right )^{2/3} .
\eea
Unitarity requires $| \mathcal{M}_{n \rightarrow n} | \leq 1$.  Setting $n = n_\text{max}$ and imposing this inequality gives a bound on the center of mass energy
\bea
\sqrt{s} \lesssim 4 \pi f \l \frac{e}{3} \log \frac{f^4}{\epsilon^4} \r^{3/2} . 
\eea
At the unitarity bound, scattering is dominated by processes involving $n_\star \sim \log \frac{f^4}{\epsilon^4}$ particles.  The energy per particle at the unitarity limit, which controls the UV cut-off,  is 
\bea
\frac{\sqrt{s}}{n_\star} \lesssim 4 \pi f \log^{1/2} \l \frac{f^4}{\epsilon^4} \r.  
\eea
It is clear that by choosing arbitrarily small $\epsilon$, the UV cutoff of the theory can be made parametrically larger than $4 \pi f$. 
The UV completion shown in Sec.~\ref{Sec: UV} will realize this limit.
A similar result holds when considering the unitarity bound from the Yukawa coupling in eq.~(\ref{eq: IR EFT}).

We  have  established that,  for small enough $\epsilon$,  our EFT still makes sense at energies above $4\pi f$, but a question lingers: what does the scale  $4\pi f$ represent physically? This question  is readily addressed by considering the $n$ dependence of $\mathcal{M}_{n \rightarrow n} $. In particular the ratio $\mathcal{M}_{n +1\rightarrow n+1}/\mathcal{M}_{n \rightarrow n}  $ measures the cost of adding two more legs to the scattering amplitude. Using  eq.~(\ref{Mn}) we find
\be
\frac{\mathcal{M}_{n +1\rightarrow n+1}}{\mathcal{M}_{n \rightarrow n} }=\left (\frac{n}{n+1}\right )^{3n-4}\left (\frac{\sqrt s}{n^{3/2}4\pi f}\right )^2\sim \left (\frac{\langle E\rangle }{n^{1/2}4\pi f}\right )^2
  \ee
where in the last step we have defined the average energy per quantum $\langle E\rangle\sim \sqrt s/n$. This result shows that $4\pi f$ is a threshold for unsuppressed multiparticle production. Notice that this does not imply strong coupling, because these processess all have small amplitudes. It is just the relative importance of processes with different $n$ that undergoes a regime change at $E\sim 4\pi f$. 

 A similar result is obtained when considering $2\to n$ processes, including in particular  $\psi \bar \psi \to n\phi$. In the $2\to n$ case the relevant quantities to compare are  the cross sections, i.e.  the squared amplitudes integrated over  phase space \footnote{The $n\to n$ amplitude of eq.~(\ref{Mn}) refers to states of unit norm. Such normalization already includes the phase space factor, see Ref.~\cite{Chang:2019vez}.}. One finds again 
\be
\frac{\sigma_{2\to n+1}}{\sigma_{2\to n}}\sim\left (\frac{\sqrt s }{n^{3/2}4\pi f}\right )^2=  \left (\frac{\langle E\rangle }{n^{1/2}4\pi f}\right )^2\,.
\ee
What these formula show is that for $n= O(1)$ the production of an additional quantum becomes unsuppressed when $\sqrt s$ becomes larger than $4\pi f$. However, because of the $n^{-3/2}$ factor, the total cross section $\sum_n \sigma_{2\to n}$ is still dominated by processes with a finite number of quanta.

Having determined the role of the scale $4\pi f$ for on-shell processes, we will now investigate virtual effects.

\subsection{Loop diagrams}

We now calculate two different corrections to the effective potential and show that these otherwise UV dominated effects are instead dominated at the  IR scale $4\pi f$.
As we  we will be  computing the UV corrections to IR physics it will be convenient to work in a background field approach and write $\phi =\phi_0+\delta\phi$, where $\phi_0$ describes the soft IR field (which in the limiting case can be taken to coincide with the  vacuum expectation value of $\phi$) , while $\delta\phi$ parametrizes the (mostly UV) quantum fluctuations.

\paragraph{Higher Harmonics.} We first consider the $\epsilon^8$ correction to the effective potential that give rise to a cosine with doubled frequency of the form
\bea
V_\text{eff} = \Delta_8 \cos^2 \l \frac{\phi_0}{f} \r \rightarrow \Delta_8 \frac{\cos \l \frac{2 \phi_0}{f} \r}{2}
\label{delta8def}
\eea
Before presenting the computation a  technical remark is in order. The 1-loop part of the Coleman Weinberg effective potential features an IR logarithm regulated by 
the mass of the scalar $m^2(\phi_0)\equiv (\epsilon^4/f^2) \cos\phi_0/f$. A proper treatment requires separating  the mass term from the potential, which slightly complicate things. As our main point concerns  the fate of the UV divergences, we found it more convenient to regulate the IR simply by adding  to the action of the quantum fluctuation a background field independent mass term $m^2\delta\phi^2/2$. The results then differs from the correct one by a finite threshold correction \footnote{This  mostly amounts to, but does not  coincide with,  the replacement $m^2\to (\epsilon^4/f^2) \cos\phi_0/f$ in the final result.}. We will not bother to do the matching computation, as it is irrelevant to our main point.

After a Wick rotation to euclidean space, $\Delta_8$ can be expressed as the path integral
\bea
\Delta_8 &=& \frac{- \epsilon^8}{2} \int d^4 x \int D \delta\phi \, e^{-S} \, \left[ e^{i \frac{\delta\phi(-x/2)}{f} + i \frac{\delta\phi(x/2)}{f} }   - e^{i \frac{{\sqrt 2}\delta\phi(0)}{f}}\l 1 - \frac{D_\phi(x)}{f^2} \r\right ] ,
\label{delta8}
\eea
where $S=\int ((\partial \delta\phi)^2+m^2\delta\phi^2)/2$ is the free action
supplemented with the IR regulator  mass $m$,
 while  $D_\phi(x)$
 is the associated  propagator.  As one can check by a straightforward diagrammatic analysis, the term in round brackets, with a proportionality constant $e^{i {{\sqrt 2}\delta\phi(0)}/{f}}$,  subtracts the non-1PI diagrams. These are those diagrams that 
 involve only zero or one propagator between $x/2$ and $-x/2$.
Expanding the exponential in $\phi$, the terms up to 4th order in $\delta \phi$ give rise upon functional integration to the usual 1-loop Coleman-Weinberg effective potential. This includes
the standard logarithmically divergent $[{\mathrm {mass}}]^4$ term.  The path integral of the higher order terms capture  all loops connecting the two Feynman vertices. It turns out that these are both calculable and physically crucial.

Calculability is evident, as eq.~(\ref{delta8}) is just a Gaussian integral. It is convenient
to rewrite the inserted fluctuations as a delta function source $J$,
\bea
\int D \delta \phi \, e^{- S} \, e^{i \frac{\delta\phi(-x/2)}{f} + i \frac{\delta\phi(x/2)}{f} } = \int D\delta \phi \,e^{- S} e^{i \int d^4 z J(z) \delta \phi(z)} \nn \\ J(z) = \frac{1}{f} \l \delta^4(z - x/2 ) + \delta^4(z + x/2) \r .\nn
\eea
 By completing the square we then find 
\bea 
\int D\delta \phi\, e^{-S} \, e^{i \frac{\delta \phi(-x/2)}{f} + i \frac{\delta \phi(x/2)}{f} } = e^{-\frac{1}{2 f^2} \l 2 D_\phi(0) + D_\phi(x) + D_\phi(-x) \r} = e^{-\frac{D_\phi(0)}{f^2}} e^{- \frac{m xK_1(m x)}{4 \pi^2 f^2 x^2}} ,
\label{unsubtracted}
\eea
where $K_1$ is the modified Bessel function. As  $\lim_{t\to 0}tK_1(t)=1$,  for $mx\ll 1$ we simply have
\be
 D_\phi(x)\Big\vert_{xm\ll1}= \frac{m xK_1(m x)}{4 \pi^2  x^2}\Big\vert_{xm\ll1}=\frac{1}{4\pi^2 x^2}\,.
 \label{UVD}
\ee
For the non-1PI subtraction part we similarly find
\be
\int D \delta\phi \, e^{-S} \, e^{i \frac{{\sqrt 2}\delta\phi(0)}{f}}\l 1 - \frac{D_\phi(x)}{f^2} \r=e^{-\frac{D_\phi(0)}{f^2}}\l 1- \frac{D_\phi(x)}{f^2} \r
\ee
which precisely corresponds to the expansion of eq.~(\ref{unsubtracted}) in $D_\phi(x)$ up to linear order,  keeping all orders in $D_\phi(0)$.

\begin{figure}[t]
\centering
\includegraphics[width=.5\linewidth]{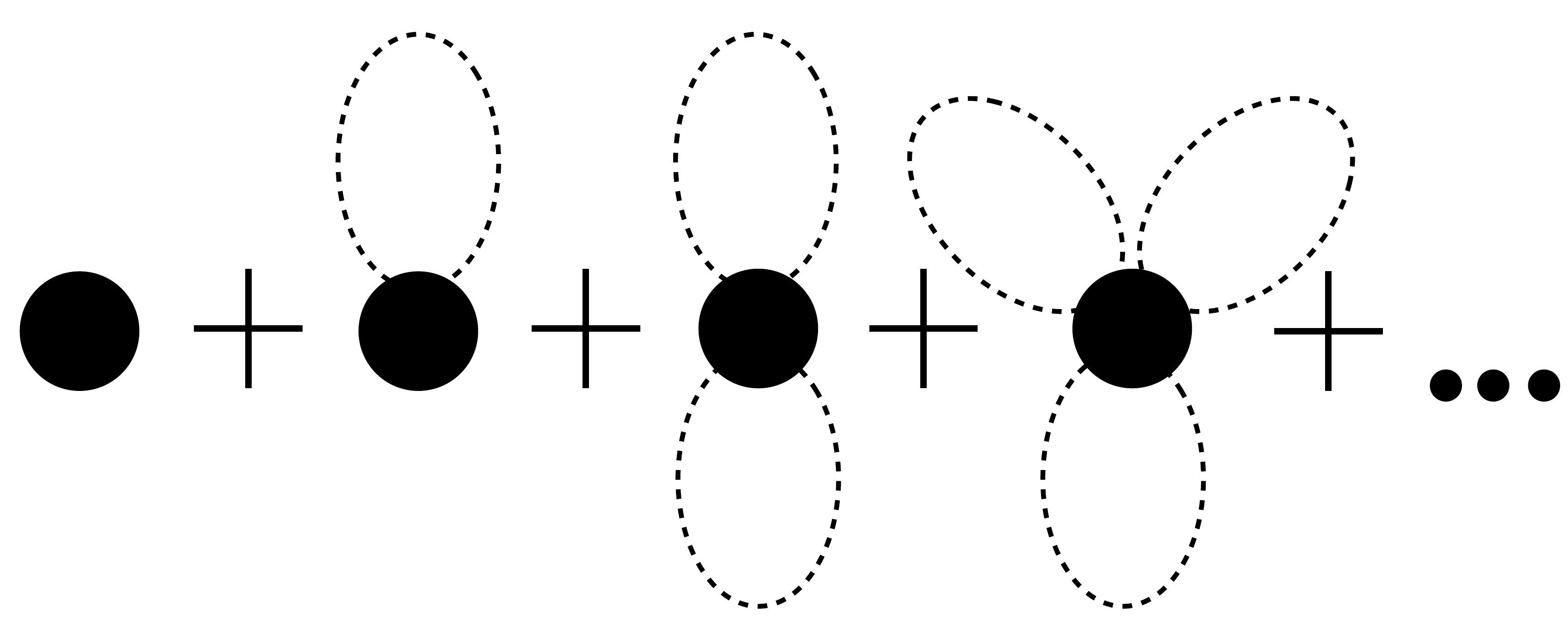}
\caption{The leading order (divergent) corrections to $\epsilon^4$.  The tree-level, 1-loop quadratic divergence, 2-loop (quadratic divergence)$^2$, and so on all sum together to give an exponential suppression of the form shown in eq.~(\ref{epsobsEFT}).  The naive expectation of divergences increasing $\epsilon^4$ is subverted and instead the tree level combined with the divergent corrections all sum into an exponential suppression.}
\label{Fig: loop}
\end{figure}

The $e^{-D_\phi(0)/f^2}$  factor represents a multiplicative renormalization of the coupling $\epsilon^4$. It can be absorbed in the  definition of the coupling observed at low energy
\be
\epsilon^4_\text{obs} = \epsilon^4 e^{-\frac{D_\phi(0)}{2 f^2}}\,.
\label{epsobsEFT}
\ee Notice that $D_\phi(0)$ coincides diagrammatically with the 1-loop tadpole, so that the exponential factor 
results from a resummation of multi-tadpole diagrams as shown in fig.~\ref{Fig: loop}.
As  $D_\phi(0)\sim \Lambda_{UV}^2/16\pi^2$, the terms in the series correspond to  the power divergences of  fixed order  perturbation theory. Remarkably, however, the resummation of the series turns the power enhancement into an exponential suppression. 
This phenomenon in  the renormalization of the $O(\epsilon^4)$ term in the potential is a prelude of what happens at $O(\epsilon^8)$.

Combining all of our results, the $O(\epsilon^8)$ correction to the effective potential reads \bea
\delta V_\text{eff} &=& - \frac{\cos \l  \frac{2 \phi_0}{f} \r}{2}  \frac{\epsilon^8_\text{obs}}{2}  \int d^4 x \l e^{- \frac{m x K_1(m x)}{4 \pi^2 f^2 x^2}} -1 + \frac{m xK_1(m x)}{4 \pi^2 f^2 x^2} \r  \nn \\
&\approx& - \frac{\cos \l  \frac{2 \phi_0}{f} \r}{2}  \frac{1}{64 \pi^2} \l \frac{\epsilon^4_\text{obs}}{f^2} \r^2 \log \l \frac{4 \pi^2 f^2}{m^2} \r.
\label{IRcos2}
\eea
Notice that, in view of eq.~(\ref{UVD}), the integral of each term in brackets is independently UV convergent. IR convergence is instead guaranteed both by mutual compensation and by the finite mass $m$.
In the last line, we have taken the small mass and large $f$ limit to isolate the logarithmically enhanced piece.
Note that the usual logarithmic UV divergence of the Coleman Weinberg potential has been cut-off at the physical scale $2 \pi f$.  As promised, the erstwhile divergent integral has been rendered finite. A glance at eq.~(\ref{delta8}) allows us to trace back the origin of this phenomenon:  two same charge operators  $e^{i\phi/f}$ inserted within a distance $x\lsim 1/(2\pi f)$ mutually cause large and ``disordered" quantum fluctuations in their  exponents that suppress their average to $e^{-1/(2\pi f x)^2}$.

\paragraph{Loops of fermions.}

Our next example is the $y^2$ correction to the effective potential.  The calculation of this contribution proceeds in much the same manner as the previous calculation. In fact the computation is simpler because this contribution does not suffer from infrared divergences in the massless limit and we can happily work with massless $\phi$ and $\psi$.

After Wick rotation the result can be written as
\bea
\label{deltaepsIR}
\delta V_\text{eff} &=& \frac{- y^2 f^2}{2} \int d^4 x \, \text{Tr} \l D_\psi(x) D_\psi(-x) \r \, \int D \delta\phi  \, e^{i \l  \delta\phi(\frac{x}{2}) +\delta\phi(-\frac{x}{2}) \r /2 f} \, e^{- S} \, \cos \l \frac{\phi_0}{f} \r  \nn \\
&=& \frac{y_\text{obs}^2f^2}{2 \pi^2} \int \frac{d x^2}{x^4} \, e^{- \frac{1}{8 f^2}\l  D_\phi(x) + D_\phi(-x) \r} \, \cos \l \frac{\phi_0}{f} \r = 8 y_\text{obs}^2 f^4 \cos \l \frac{\phi_0}{f} \r ,
\eea
where  the massless fermion propagator $D_\psi(x)$ gave  $\text{Tr} \l D_\psi(x) D_\psi(-x)\r =-1/(\pi^4 x^6)$; moreover  we used eq.~(\ref{UVD})  and similarly to before defined
\be
y_\text{obs} = y e^{-\frac{D_\phi(0)}{8 f^2}}\,.
\label{yobsEFT}
\ee
This time, the standard quadratically divergent loop integral has been rendered finite with the UV cutoff instead replaced by $\sim 4 \pi f$ \footnote{ The factor of $2$ increase in the UV cut-off with respect to the previous case stems from the $1/2$ in the phase factor of the  Yukawa interaction, see eq.~(\ref{eq: IR EFT}).}.

Notice, amusingly,  that in the self-consistent situation where eq.~(\ref{deltaepsIR}) represents the leading contribution
to the scalar potential, there is a sharp relation between the masses  $m_\psi$ and $m_\phi$ of respectively fermion and  boson. The minimum of the potential in eq.~(\ref{deltaepsIR}) is at $\phi/f=\pi$ which breaks the discrete chiral symmetry protecting the fermion mass. Expanding around the minimum we find $m_\phi=2m_\psi$. The fact that $m_\phi$  is not suppressed with respect to  $m_\psi$, even though it is generated from loops involving the fermion mass interaction is because quantum fluctuations of $e^{i\phi/f}$ at the scale $4\pi f$ are unsuppressed \footnote{For instance, in a strongly coupled composite Higgs model where the strong scale is $\sim 4\pi v_F\sim 2$ TeV coincides with the UV-cut-off, one expects, according to eq.~(\ref{deltamh}),  a correction $\delta m_H\sim m_t$ with no suppression.}.

\paragraph{General Operators.}

Our two  examples  are not exceptions.  The dressing of erstwhile divergences with exponentials is a general phenomenon.  If the scalar couples to an operator as
\bea
\delta \mathcal{L} = y \sin \l \frac{\phi}{2 f} \r \mathcal{O},
\label{intO}
\eea
then the effective potential will be corrected as
\bea
V_\text{eff} \propto y^2 \cos \l \frac{\phi}{f} \r \int \frac{d^4 x}{x^{2 \Delta_\mathcal{O}}} \, e^{- \frac{1}{16 \pi^2 f^2 x^2}} \sim  y^2 \cos \l \frac{\phi}{f} \r \, \Gamma(\Delta_\mathcal{O}-2)  \, \l 4 \pi f \r^{2 \Delta_\mathcal{O} - 4}
\eea
where $\Delta_\mathcal{O}$ is the dimension of the operator $\mathcal{O}$.  The integral is dominated at a scale $4 \pi f \sqrt{\Delta_\mathcal{O}}$.  Only for large enough ${\Delta_\mathcal{O}}\gg 1$  can this  effective cut-off approach  the physical UV cutoff of the theory.

\section{UV completion} \label{Sec: UV}

The previous section presented an EFT endowed with a physical scale $4\pi f$ within its perturbative domain of validity that controls the onset of  a  dynamical  regime where multiparticle states
dominate both real and virtual processes.  In this section, we present a UV completion of our EFT. That  will show both its robustness and the microphysics features 
that are necessary to give rise to it. In particular we will find that the request $4\pi f\ll \Lambda_{UV}$  requires the presence  of a large discrete symmetry, ${\mathbb Z}_N$ with $N\gg 1$.

The basis of our construction is the theory of a complex scalar $\Phi$ and of a Dirac fermion $\psi$, endowed with a ${\mathbb Z}_2\rtimes {\mathbb Z}_N$ symmetry under which 
\bea{\mathbb Z}_2&:&\quad\Phi\rightarrow \Phi^\dagger\,,\qquad\qquad  \qquad\bar\psi \psi\rightarrow - \bar\psi \psi\\
{\mathbb Z}_N&:&\quad \Phi \rightarrow e^{\frac{2 \pi k i}{N}} \Phi\,,\qquad\qquad\,\,\bar \psi \psi \rightarrow (-1)^k \bar\psi \psi
\quad \quad k=1.\dots,N-1
\eea

The Lagrangian containing the lowest dimension operators consistent with ${\mathbb Z}_2\rtimes {\mathbb Z}_N$ contains, besides the kinetic terms, the potential 
\bea \label{Eq: fullUV}
V = -m_\Phi^2 | \Phi |^2 + \frac{\lambda_\Phi}{4} | \Phi |^4 - \lambda_N \l \Phi^N +\Phi^{\dagger, N}\r + iY \l \Phi^{N/2} - \Phi^{\dagger, N/2} \r \bar \psi \psi\, .
\eea
Notice that the terms proportional to $\lambda_N$ and $Y$ serve the role of explicitly breaking the $U(1)$ symmetry $\Phi\to e^{i\alpha}\Phi$ down to ${\mathbb Z}_N$.


The negative mass term in the potential forces $\Phi$ to acquire an expectation value. This would give rise to a Goldstone boson in the limit of an exact $U(1)$. As we want to maintain a light pseudo-NG boson to match the EFT of the  previous section, we will work under the assumption that  the couplings $\lambda_N$ and $Y$, which explictly break the $U(1)$, can be treated  as {\it small} perturbations. A small $Y$ also ensures, under all circumstances, the lightness of the fermion $\psi$. We can then expand around the minimum as
\bea
\la \Phi \ra = \frac{F + \rho}{\sqrt{2}} e^{i \phi/F}\,,
\eea
where at lowest order in $\lambda_N$ the expectation value $F$ and the mass of of the radial mode $m_\rho$ are given by
\be
F^2  = \frac{4 m_\Phi^2}{\lambda_\Phi}\,, \qquad\qquad m^2_\rho = 2 m_\Phi^2\,.
\label{UVthresh}
\ee

At energies below $m_\rho$ the radial mode $\rho$ can be integrated out. At the lowest order in the derivative expansion and up to linear order in $\lambda_N$ and $Y$, the  resulting  effective Lagrangian for  $\phi$ and $\psi$ matches eq.~(\ref{eq: IR EFT}) with 
\be f=\frac{F}{N}\,,\qquad \epsilon^4 = 2\lambda_N \left (\frac{F}{\sqrt2}\right )^N\,,\qquad y= Y N \left (\frac{F}{\sqrt 2}\right )^{N/2-1}\,.
\label{match}
\ee
Higher orders in $\lambda_N$ and $Y$ involve, in particular,  contributions to higher harmonics. In order for our story  to make sense,  those UV induced  contributions, at tree level and beyond, should be subdominant to the loop induced IR ones, which we computed in the previous section. Our goal now  is   to show that this the case.  Indeed the zeroth order request for our story to make sense is that the EFT UV cut-off $m_\rho$ be parametrically larger than the scale $4\pi f$ at which power divergences are cut-off in the EFT. Using eqs.~(\ref{UVthresh},\ref{match}) this constraint reads
\be
\frac{m_\rho^2}{16\pi^2 f^2}\sim \frac{\lambda_\Phi N^2}{16\pi^2}\gg 1 . \label{laNsquare}
\ee
%
In a  weakly coupled  model, i.e.
with $\lambda_\Phi/16\pi^2\ll 1$, eq.~(\ref{laNsquare}) necessarily requires $N\gg 1$, corresponding to a large charge  interaction\footnote{In this situation standard perturbation theory works reliably for  $\lambda_\Phi N/16\pi^2\ll 1$, while for $\lambda_\Phi N/16\pi^2\gsim 1$, one must instead employ a slightly more involved but equally reliable,   semiclassical method \cite{Badel:2019oxl}. The scaling $\lambda_\Phi N^2=$ fixed for $N\to \infty$ was,  to our knowledge, first considered in \cite{Arias-Tamargo:2019xld}.}. Consistent with eq.~(\ref{laNsquare}) and also to simplify our computations, we  find it convenient to scale our parameters as $N\to \infty $ so that 
\be
\gamma=\frac{\lambda_\Phi N^2}{16\pi^2} ={\mathrm{fixed}}\gg 1\,,\qquad m_\rho={\mathrm{fixed}}\,,\qquad f=\frac{F}{N}=
{\mathrm{fixed}}\,,
\label{scaling}
\ee 
where, by $m_\rho^2 \sim \lambda_\Phi F^2$, only two of the above relations are independent. Notice also that the first relation implies that ${\lambda_\Phi N}/{16\pi^2}$, which controls the  Feynman diagram expansion  at  large charge, goes to zero as $N\to \infty$.

%
%
%
%
%
%
%

Consider now the parameters $Y$ and $\lambda_N$. The classical dimensions of the corresponding operators 
scales like $N$ and go to infinity in the scaling limit depicted above. Their 1-loop anomalous dimension is proportional to $\gamma$ in eq.~(\ref{scaling}), while higher-loop contributions are instead controlled by 
powers of $\eta={\lambda_\Phi N}/{16\pi^2}$. As  in our scaling limit $\gamma=$ fixed and $\eta\to 0$, we will be allowed to limit our analysis to a sublass of (suitably resummed) 1-loop effects.

To stay on the safe side we want to treat $Y$ and $\lambda_N$ as small perturbations.  Considering for definiteness $\lambda_N$, the constraints  $\delta\langle \Phi\rangle/\langle \Phi\rangle\lsim 1$ and $\delta m_\rho/m_\rho\lsim 1$ read respectively 
\be
\epsilon^4=2\lambda_N (F/\sqrt 2)^N\lsim F^2 m_\rho^2 N^{-1}\, , \qquad \epsilon^4=2\lambda_N (F/\sqrt 2)^N\lsim F^2 m_\rho^2 N^{-2}\, .
\ee
The bound gets stronger when going from the 1- to the 2-point function, and indeed the higher point functions give increasingly stronger bounds.  The strongest constraint comes in the end from amplitudes with $n\sim N$ legs.
The computation of these amplitudes, at least close to threshold, was addressed long ago, focussing on processes of the form 1 virtual $\to n-1$ real.  At tree level, the matrix element 
$\langle n-1| \rho|0\rangle$, with $\langle n-1|$ the bra of $n-1$ $\rho$-quanta,  is unaffected by the pseudo-NG $\phi$, and, using refs.~\cite{Brown,Voloshin} we find
\be
{\cal A}_{1\to n-1}\equiv \langle n-1| \rho|0\rangle = (n-1)! \left (\frac{1}{2 f}\right )^{n-2} .
\ee
On the other hand, $\lambda_N$ gives a correction
\be
\delta {\cal A}_{1\to n-1}= \frac{2\lambda_N (F/\sqrt 2)^N}{(F)^n}\frac{1}{m_\rho^2} \frac{N!}{(N-n)!}\frac{1}{n(n-2)}
\ee
where the factor $1/(m_\rho^2 n(n-2))$ is the propagator of the incoming off-shell $\rho$-quantum that disintegrates into $n-1$ real quanta.
Cautiously requesting $\delta {\cal A}_{1\to n-1}\lsim {\cal A}_{1\to n-1}$, the strongest constraint is 
given by $n\sim N/2$ and reads
\be
\epsilon^4 = 2\lambda_N (F/\sqrt 2)^N \lsim Nm_\rho^2 F^2 2^{3-\frac{3}{2}N}\sim F^2 m_\rho^2 \,e^{-\frac{3N}{2}\ln {2}}
\ee
corresponding to an exponential suppression with respect to vacuum energy scale of the original complex scalar.
Notice that this exponential suppression guaranteed that the order of magnitude of $\ln \epsilon^4$ is not significantly affected by threshold corrections at the RG matching scale $\mu_{RG}\sim m_\rho$, as the anomalous dimension $\gamma$ is large but still $\ll O(N)$.

A similar exponential suppression  can be derived for $Y$, and $y$, by considering the 1-loop contribution to the same purely bosonic process.

%

\paragraph{The higher harmonics in the full UV theory.}

We want to repeat the computations done in Sec.~\ref{Sec: EFT}, but this time within  the UV completion. As already stated, our main goal is  to test the robustness of our conclusion that power divergent loops are saturated below the UV cut-off $m_\rho$ of the low energy EFT.

The main novelty in the full theory is the propagating radial mode $\rho$.
Eq.~(\ref{scaling}) offers a great simplification, as it implies $\lambda_\Phi\to 0$:  the quantum effects of  $\lambda_\Phi$ vanish, if not enhanced by two powers of $N$. That means  we can safely truncate to quadratic order the kinetic and  $\lambda_\Phi$ parts of the Lagrangian, while keeping higher powers of $\rho, \phi$ only in the terms involving $O(N)$ legs \footnote{One should not be confused by the fact that $\lambda_N$ and $Y$ also go to zero, and exponentially so, when $N\to \infty$. We keep the effects of these terms as they are the leading ones  involving the breaking of $U(1)$ to ${\mathbb Z}_N$ and producing a potential for $\phi$, no matter how suppressed.}. The fact that $\rho/F$ can be treated as an infinitesimal quantity also gives us the right  to ``exponentiate" $\rho$ when taking $O(N)$ powers of $\Phi$
\be
\Phi^N=\left (\frac{F}{\sqrt 2}\right )^N \left (1+\frac{\rho}{Nf}\right )^N e^{i\frac{\phi}{f}} \simeq \left (\frac{F}{\sqrt 2}\right )^N
e^{\frac{\rho+i\phi}{f}} .
\label{expapprox}
\ee
The above step, while intuitive, may seem a bit cavalier. We shall later come back and check that it is justified in the scaling limit of  eq.~(\ref{scaling}).

In view of the above comments, the UV dynamics is described  by the Lagrangian
\bea
\label{UVQFT}
\mathcal{L}& =& \frac{1}{2} \l \partial \rho \r^2-\frac{1}{2} m_\rho^2 \rho^2+ \frac{1}{2} \l \partial \phi \r^2 + i \overline \psi\slashed\partial \psi \nn\\
&+&\frac{1}{2}\epsilon^4 \l e^{\frac{\rho+i\phi}{f}}+{\mathrm{h.c.}} \r - \frac{i}{\sqrt{2}} y f \l e^{\frac{\rho+i\phi}{2f}}-{\mathrm{h.c.}} \r\overline \psi \psi.
\eea
At tree level, the matching to the IR theory of eq.~(\ref{eq: IR EFT}) simply amounts to setting $\rho=0$.
To compute  quantum corrections to the $\phi$ potential, like before we decompose $\phi=\phi_0+\delta\phi$ and integrate out   $\rho$,  $\delta\phi$ and $\psi$. 

Let us consider first the multiplicative renormalization of $\epsilon^4$ and $y$. This coincides with  the multiplicative renormalization of respectively $\lambda_N$ and $Y$. The UV divergent part of these corrections is therefore related to the anomalous dimensions of the corresponding operators. Thanks to the structure of eq.~(\ref{UVQFT}) the observed (i.e. at low momenta)  couplings are again determined  by  a strightforward gaussian integral, and read
\bea
\epsilon^4_{obs}=\epsilon^4  e^{\frac{1}{2 f^2} \l D_\rho(0) - D_\phi(0) \r }\,,\qquad y_\text{obs} = y e^{\frac{1}{8 f^2} \l D_\rho(0) - D_\phi(0) \r } .
\eea
Notice that the real and purely imaginary factors with which respectively $\rho$ and $\phi$ appear in eq.~(\ref{expapprox}), translate into opposite signs for the propagators in the exponents. This ensures the exact cancellation of the leading quadratic divergence, which instead appeared in the EFT computation (see eqs.~(\ref{epsobsEFT}, \ref{yobsEFT})). The residual logarithmic divergence coincides with the anomalous dimension of the corresponding operators. Indeed, by eqs.~(\ref{UVthresh},\ref{match}),  one has
\be
\frac{1}{2 f^2} \l D_\rho(0) - D_\phi(0) \r = \frac{1}{2 f^2}\int \frac{d^4 p}{(2\pi)^4} \frac{-m_\rho^2}{p^2(p^2+m_\rho^2)}\simeq -\frac{\lambda_\Phi N^2}{32\pi^2}\ln \frac{\Lambda}{m_\rho}\equiv -\gamma_{N}\ln \frac{\Lambda}{m_\rho}
\ee
properly reproducing the $N\gg 1$ limit of the $\Phi^N$ anomalous dimension  $\gamma_{N}$  (see e.g. \cite{Badel:2019oxl}). 
The exponent in $y_\text{obs}$ is similarly controlled by the $\Phi^{N/2}$ anomalous dimension $\gamma_{N/2}$.

Next we calculate the $y^2$ correction  to $\epsilon^4$, as it is emblematically similar to the top correction to the Higgs mass (the computation of the other corrections, like the $\epsilon^8$ one, goes however along the same lines). The result parallels eq.~(\ref{deltaepsIR}) with the extra contribution from $\rho$ exchange. Working in euclidean space and indicating the free euclidean quadratic action for $\rho$ and $\delta\phi$ simply by $S$, we thus find
%
\bea
\delta\epsilon_{obs}^4&=& - \frac{y^2 f^2}{2} \int d^4 x \, \text{Tr} \l D_\psi(x) D_\psi(-x) \r \, \int D  \rho \, D \delta \phi \,  \, e^{\l  \rho(\frac{x}{2}) + i \delta \phi(\frac{x}{2}) +  \rho(-\frac{x}{2}) + i \delta \phi(-\frac{x}{2}) \r /2 f} \, e^{- S} \nn \\
&=& \frac{y^2f^2}{2 \pi^2} \int \frac{d x^2}{x^4} \, \int D  \rho \, D \delta \phi \, \, e^{\l  \rho(\frac{x}{2}) + i \delta \phi(\frac{x}{2}) + \rho(-\frac{x}{2}) + i \delta \phi(-\frac{x}{2}) \r /2 f} \, e^{- S} \nn \\
&=& \frac{y^2f^2}{2 \pi^2} \int \frac{d x^2}{x^4} e^{\frac{1}{8 f^2} \l 2 D_\rho(0) - 2 D_\phi(0) + D_\rho(x) - D_\phi(x) + D_\rho(-x) - D_\phi(-x) \r} \nn\\
&=& \frac{y_{obs}^2f^2}{2 \pi^2} \int \frac{d x^2}{x^4} e^{\frac{1}{8 f^2} \l  D_\rho(x) - D_\phi(x) + D_\rho(-x) - D_\phi(-x) \r}\equiv \frac{y_{obs}^2f^2}{2 \pi^2} \int \frac{d x^2}{x^4} e^{-\frac{\gamma_N}{2} G(x m_\rho)} .
\label{deltaepsUV}
\eea
As expected this matches the EFT result in eq.~(\ref{deltaepsIR}), apart from the $\rho$ contribution in the exponent. $D_\rho(x)$ is however exponentially suppressed for $m_\rho x\gg 1$ so that, in the EFT domain, eq.~(\ref{deltaepsUV})  coincides with eq.~(\ref{deltaepsIR}). It is however interesting to study the behaviour of the integrand for arbitrary $m_\rho x$.  For that purpose in the last line of eq.~(\ref{deltaepsUV}) we have conveniently  expressed the exponent  as
\bea \label{Eq: exponent}
\frac{1}{8 f^2} \big ( D_\rho(x) - D_\phi(x) + D_\rho(-x) - D_\phi(-x) \big ) &=&-\frac{\gamma_N}{2} G(m_\rho x)\\
G(z) &\equiv&   \frac{2 -2z\,  K_1(z)}{z^2}\, .
\eea
Now, crucially, $G(z)$ is  both positive definite and monotonically decreasing. This guarantees that the  integral in eq.~(\ref{deltaepsUV}) is indeed saturated at length scales $\sim 1/4\pi f$ where the exponent is $O(1)$, while the contribution of shorter scales, in particular the UV/IR matching scale $1/m_\rho$, is suppressed. Even if the region $z=xm_\rho \ll 1$ is subdominant in the integral, it is interesting to study the behaviour of the integrand there. Indeed in that region one has the asymptotic behaviour $G(z) \approx -  \log z $, so that the exponential factor becomes
\be
e^{-\frac{\gamma_N}{2} G(xm_\rho)}\propto  \l x m_\rho \r^{\frac{\gamma_N}{2}}\equiv  \l x m_\rho \r^{\gamma_N-2\gamma_{N/2}}
\label{asymptG}
\ee
where in the last step we  used $\gamma_N=\lambda_\Phi N^2/32\pi^2$ to express the exponent as the difference of the anomalous dimensions of respectively ${\cal O}_N\equiv \Phi^N$ and ${\cal O}_{N/2}\equiv \Phi^{N/2}$. The above result matches the OPE  in the  far UV regime 
where our theory (with $\lambda_N=Y=0$) is at lowest order conformally invariant. Indeed the UV tail of our computation, in the language of conformal perturbation theory,
is  controlled by the OPE
\be
{\cal O}_{N/2}(x){\cal O}_{N/2}(-x) =C x^{\Delta_N-2\Delta_{N/2}} {\cal O}_N(0)= C x^{\gamma_N-2\gamma_{N/2}} {\cal O}_N(0)\, .
\label{OPE}
\ee
which by use of $\Delta_N= N+\gamma_N$  perfectly matches eq.~(\ref{asymptG}). The convexity of the operator dimension   as a function of charge \cite{Badel:2019oxl} guarantees $\Delta_N-2\Delta_{N/2}>0$ and thus the UV convergence of the correction. 

\begin{figure}[t]
\centering
\includegraphics[width=.6\linewidth]{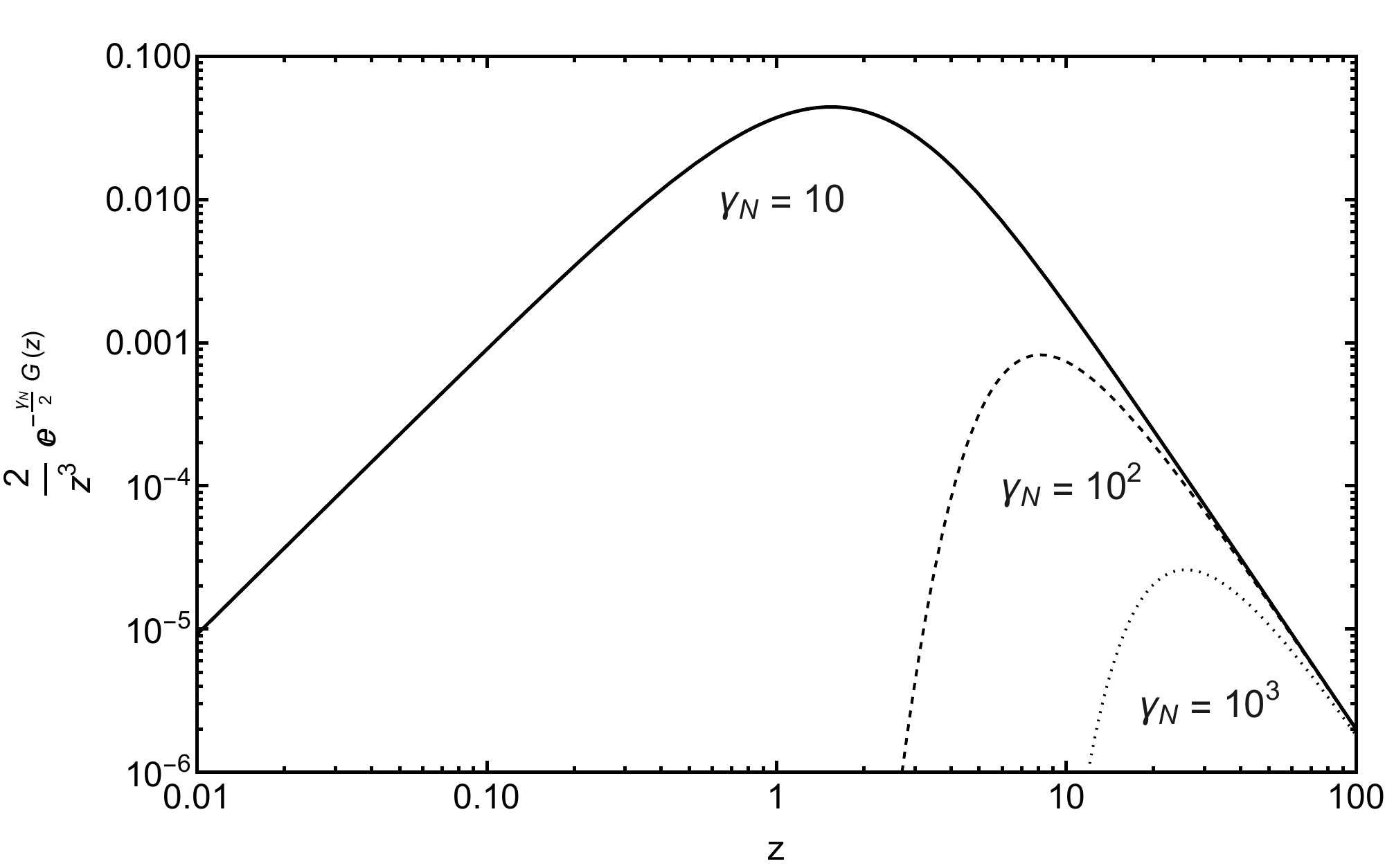}
\caption{The contribution of a loop of fermions as a function of distance to the mass of $\phi$, i.e. the integrand of eq.~(\ref{Eq: mass}) as a function of $z = m_\rho x$ where $x$ is the distance between the two vertices.  The solid line is when $\G = 10$, the dashed line is when $\G = 10^2$ while the dotted line is when $\G = 10^3$.  At low energies, large z, the standard quadratic growth is seen.  At a scale $z^2 \sim \G$, the fermion loop saturates and begins to be suppressed indicating the the fermion loop has its structure changed at scales well below the mass of the $\rho$.  This demonstrates that the structure of divergences changes at a scale not associated with the mass of the $\rho$, $z=1$, and takes the initially divergent fermion loop and renders it finite.}
\label{Fig: integrand}
\end{figure}

It is convenient to discuss the integral in eq.~(\ref{deltaepsUV}) by using  the dimensionless coordinate $z=xm_\rho$.  It is also  suggestive to translate $\delta\epsilon^4$ into a correction $\delta m_\phi^2$ to the mass term for $\phi$, which then reads
\bea
\label{Eq: mass}
\delta m_\phi^2 = -\frac{\delta\epsilon^4}{f^2}= - \frac{y_{obs}^2m_\rho^2}{2 \pi^2}  \int \frac{d z^2}{z^4} e^{- \frac{\gamma_N}{2} \, G(z)}\,.
\eea
The prefactor of the dimensionless integral represents the  estimate of $\delta m_\phi^2$, based on a  naive application of dimensional analysis and selection rules. Our construction however features another control parameter, $\gamma_N$, which given $G(z\sim 1)=O(1)$, exponentially suppresses the integrand in the naively dominant threshold region $z\sim 1$. The integral is thus instead dominated at $z\sim \gamma_N$ (i.e. $x\sim 1/4\pi f$) giving a $1/\gamma_N$ suppression with respect to the naive result
\be \delta m_\phi^2 \sim-\frac{y_{obs}^2(4 \pi f)^2}{2 \pi^2} =  -\frac{y_{obs}^2m_\rho^2}{2 \pi^2}\times \frac{1}{\gamma_N}, .\ee
We should stress that it is essential for this result that $G(z\sim 1)=O(1)$. When comparing to the OPE in eq.~(\ref{OPE}) this corresponds to a Wilson coefficient $C\sim m_\rho^{\gamma_N-2\gamma_{N/2}}e^{-\gamma_N \times O(1)}$, i.e.  exponentially suppressed with respect to its naivest estimate. This exponential suppression ensures that the $dx$ integral is dominated at  even longer distances than guaranteed by the CFT regime in eq.~(\ref{OPE}).  Our computation shows that the presence of a large charge $N$  ``extends the UV suppressing arm'' of the  OPE beyond the scale invariant regime. Fig.~\ref{Fig: integrand} offers a graphical representation of these results.

It is easy and instructive to study how eq.~(\ref{Eq: mass}) changes when the fermion bilinear is replaced by the interaction in eq.~(\ref{intO}). Without paying attention to $O(1)$ factors, in that case one has
\bea
\label{massO}
\delta m_\phi^2 \propto  \frac{y_{obs}^2 }{2 \pi^2} \, m_\rho^{2\Delta_{\mathcal O} - 4} \int \frac{d z^2}{z^{2\Delta_{\mathcal O}-2} }e^{- \frac{\gamma_N}{2} \, G(z)}\,.
\eea
For $\Delta_{\mathcal O}=O(1)$ the numerical integral will be similarly saturated at large $z$, ending up in a suppression with respect to the naive estimate in the prefactor. However for $\Delta_{\mathcal O}\gg 1$ the integral will be saturated closer to $z\sim 1$ reducing the suppression. A simple saddle point estimate of the integral indeed gives  $(\Delta_{\mathcal O}/\gamma_N)^{2\Delta_{\mathcal O}-4}$: a suppression persists only as long as $\Delta_{\mathcal O}\ll\gamma_N$.

\paragraph{Scales in the QFT and exponentiation.}

Working at finite but large $N$, the QFT under consideration has many scales.  Going from long distance to short distances, the first scale of import is  $f = F/N$ where multiparticle processes set in and where  some loop integrals are dominated.  Next, there is $m_\rho$, the physical threshold acting as UV cut-off of the low energy EFT.  Then  is the scale $F/\sqrt{N} \equiv 1/x_2$, at which  the linearization of $\rho$ interactions with finite number of legs and  their $N$-leg exponentiation fail.  Finally there is the scale $\sqrt{N} F \equiv 1/x_1$ where complete symmetry restoration occurs. In the $N\to \infty$ scaling of eq.~(\ref{scaling}) only $f$ and $m_\rho$ stay finite, while $x_{1,2}\to 0$. The phenomena pertaining these latter scales thus do not concern us. This means, in particular, that exponentiation of the radial mode happily applies at all scales. Let us see that.

To see the scale $x_2$, consider the exponentiation
\bea
\l 1 + \frac{\rho}{F} \r^N \rightarrow e^{N \rho/F} .
\eea
The Taylor series of an exponential $e^c$ is dominated by the $c^\text{th}$ terms.  As we are exponentiating the propagators, $D_\rho(x)\sim 1/(4\pi^2 x^2)$, requiring that the exponent is smaller than $N$ gives
\bea
\frac{N^2}{F^2 x^2} \lesssim N \qquad \Rightarrow \qquad x_2 = \frac{\sqrt{N}}{F}
\eea
only for $x \gtrsim x_2$ is the exponentiation trick that we utilize valid.  The above constraint also coincides with the linearity request $\rho/F<1$ for  the configurations that dominate the path integral. When performing the gaussian integral we have $\rho/F \sim N/F^2x^2$, which again translates into $x> x_2$. As $x_2\propto 1/\sqrt N$ in the scaling of eq.~(\ref{scaling}) we conclude that the linearization and exponentiation expressed by 
eq.~(\ref{UVQFT}) are exact for $N\to \infty$.

Finally, let us consider the scale where complete symmetry restoration occurs.  For that purpose we can consider for instance the correlator  $\Phi^N(x) \Phi^{\dagger,N}(0)$ and study where the first correction proportional to $F^2$ becomes comparable to the leading short distance result. One has
\be
\langle \l \frac{F+\rho+i\phi}{\sqrt 2}\r ^N(x)\,\l \frac{F+\rho-i\phi}{\sqrt 2}\r^N(0)\rangle=N!\left [D(x)^N+\frac{N}{2}F^2 D(x)^{N-1}+\dots\right ]
\ee
  Requiring that the first term is more important gives
\bea
\frac{1}{x^2} \gtrsim N F^2 \qquad \Rightarrow \qquad x_1 = \frac{1}{\sqrt{N} F}
\eea
At length scales smaller than $x_1$ the vev of $\Phi$ can be completely neglected, indicating complete symmetry restoration.

\paragraph{Non-IR dominated loop integrals.}

We must point out, in order to avoid confusion, that the dominance of loop integrals  at the  scale $4\pi f$ is not a universal feature.  That interesting  phenomenon only occurs when considering insertions carrying $U(1)$ charges of the same sign. That is already clear when considering the UV regime where our QFT is well approximated by a CFT. For instance, if, instead of same charge operators in eq.~(\ref{OPE}), we considered operators of opposite charge, the OPE would involve many terms singular at $x\to 0$, starting with the one  associate with the identity operator
\be
{\cal O}_{N/2}(x){\cal O}^\dagger_{N/2}(-x) =\frac{1}{x^{2\Delta_{N/2}}}+ \dots\, .
\ee
Quantum corrections would then be fully UV dominated. Relatedly, in our full theory, involving  finite $m_\rho$, the integrand
is controlled by the exponential factor
\bea
e^{\gamma_N  \frac{1 + m_\rho \, x \,  K_1(m_\rho \, x)}{m_\rho^2 \, x^2}}
\eea
which  as signalled by the positive sign in the exponent makes now the integral UV dominated.   The change of sign in the exponent gives exponential enhancement where we previously had  an exponential suppression.

%
%

%
%
%
%
%
%

\section{Conclusion} \label{Sec: conclusion}

The application of naturalness to the Higgs boson has driven the field of particle physics for many years.  
The most important aspect of this application of naturalness is the prediction of where new particles should appear.  In this article, we demonstrated that the scale predicted by dimensional analysis is not necessarily the scale where new particles appear. 

By considering a shift symmetric Yukawa coupling, we showed that the usual quadratic divergence is not present and is instead regulated by a new scale $\sim 4 \pi f$ that is parametrically smaller than the scale where new particles appear.  It is this scale that is predicted by naturalness, rather than the scale of new particles.  The physics that appears at this energy scale manifests itself as the importance of multiple loops/final states.

In light of the apparent lack of new particles at the meV scale and the TeV scale, it would be interesting if these thoughts could be applied to the cosmological constant or the Higgs boson.  Speculations and progress along these directions are left as an exercise for the reader.

\section*{Acknowledgement}
We thank Markus Luty for collaboration in the early stages of the project and  acknowledge hospitality of KITP Santa Barbara where this project was started. RR is indebted with Sasha Monin for many insightful discussions.
AH is supported by the NSF grant PHY-2210361 and by the Maryland Center for Fundamental Physics (MCFP). RR is partially supported by the Swiss National Science Foundation under contract 200020-213104 and through the National Center of Competence in
Research SwissMAP.
\appendix
\section{Higher spin symmetry selection rules}

Here we would like to briefly illustrate the role of the higher spin symmetries of free field theory  in shaping the notion of naturalness. These symmetries are best characterized by working in momentum space. It also suffices to focus on the simplest case of a  massive free scalar $\phi$, whose action can be written in spacetime and momentum  space as
\be
S=\int d^4 x \phi(x)(-\partial^2-m^2)\phi(x)=\int \frac{d^4p}{(2\pi)^4} \hat \phi (-p)(p^2-m^2)\hat \phi(p)
\ee
where $\hat \phi(-p)=\hat \phi(p)^*$. It is evident from the momentum space representation that the transformation
\be
\hat \phi(p)\to e^{i\theta(p)}\hat \phi(p) \qquad\qquad \theta(-p)=-\theta(p)
\label{free}
\ee
is a symmetry of the action. By expanding $\theta$ in a power series $\theta=a_\mu p^\mu+a_{\mu\nu\rho}p^\mu p^\nu p^\rho+\dots$ the transformation in position space reads
\be
\phi \to (1 +a_\mu\partial^\mu -a_{\mu\nu\rho}\partial^\mu\partial^\nu\partial^\rho +\dots)\phi
\ee
for which the Noether curents are  the tensor bilinears in $\phi$  with even rank $r\geq 2$  (the energy momentum tensor corresponding to $r=2$), hence the label {\textit{higher spin symmetry}}. One can easily see that the infinite symmetry eq.~(\ref{free}) ``protects'' all terms in the action other than the quadratic ones. Indeed the effective  1PI action of a generic QFT can be formally made invariant by  assigning to its  $n$-point functions   the transformation property
\be
\Gamma^{(n)}(p_1,\dots, p_n)\to \Gamma^{(n)}(p_1,\dots, p_n)e^{-i\sum_{a=1}^{a=n}\theta(p_a)}\,.
\ee
For a general $\theta$ satisfying eq.~(\ref{free}), true invariance of the action corresponds to the above transformation 
being the identity, which, for well behaved  $\Gamma^{(n)}$, can only happen for the two point function
 $\Gamma^{(2)}=\delta^{(4)}(p_1+p_2)\Gamma(p_1)$. 
This infinite symmetry thus protects all vertices with more than two legs, that is all the non-trivial interactions.

To learn  the implications of the  selection rules associated with the high spin symmetry, we can focus on 
the theory of a real scalar and a Dirac fermion with Lagrangian
\be
\mathcal{L} = \frac{1}{2} \l \partial \phi \r^2 + i \overline \psi\slashed\partial \psi -\lambda\phi^4+ y\phi\bar \psi\psi\, .
\ee
Working in momentum space, the couplings can be viewed as momentum dependent spurions,  $\lambda\Rightarrow \hat \lambda(p_1,p_2,p_3,p_4)$ and $y\Rightarrow \hat y(p_1,p_2,p_3)$,
transforming as
\bea
 \hat \lambda(p_1,p_2,p_3,p_4)&\longrightarrow & \hat \lambda(p_1,p_2,p_3,p_4)e^{-i(\theta_\phi(p_1)+\theta_\phi(p_2)+\theta_\phi(p_3)\theta_\phi(p_4))} \\
  \hat y(p_1,p_2,p_3)&\longrightarrow & \hat y(p_1,p_2,p_3) e^{-i(\theta_\phi(p_1)+\theta_\psi(p_2)+\theta_\psi(p_3))}
  \label{spurions}
  \eea
 When considering the effective action (let us say the 1PI action for definiteness) one can easily see the implications of this symmetry: the compensation of the phase rotations of the external fields, see eq.~(\ref{free}), and of the spurions, eq.~(\ref{spurions}), give rise to precisely the structures of the constructable  Feynman diagrams.
Use of these selection rules therefore does not teach us anything that we could not learn by drawing Feynman graphs. Still we find it conceptually important that the allowed combinations are controlled by an infinite number of symmetries.

The conceptual relevance is evidenced when considering naturalness issues, which are too often mistakenly viewed as laking sufficient formal ground. The scalar mass in the above theory is a good example to illustrate that.
What we must consider are the possible contributions to the 2-point function $\Gamma(p_1,p_2)$ for $\phi$. Assuming  the above theory is endowed with a physical UV cut-off $\luv$, the combined selection rules of dilation ($\equiv$ dimensional analysis) and higher spin symmetry are easily seen to allow the following corrections from respectively $\lambda$ and $y$
\be
 \delta \Gamma(p_1,p_2)\propto \hat \lambda(p_1,p_2, k, -k) \luv^2\,,\qquad  \delta \Gamma(p_1,p_2)\propto \hat y(p_1,k_1,k_2)\hat y(p_2,-k_1,-k_2) \luv^2\,.
  \ee
Replacing then the spurions with their physical values, i.e.  $\hat \lambda\to (2\pi)^4\delta(p_1+\dots+p_4) \lambda$ and
$\hat y\to (2\pi)^4\delta(p_1+p_2+p_3) y$, we see that the above structures are precisely those encountered when estimating the 1-loop corrections to the scalar mass. This argument clarifies  the ubiquity of the structure $(\mathrm{coupling})^2(\mathrm{cutoff})^2$ for mass corrections, which was exemplified in the introduction.

Where does the model discussed in this paper stand in the face of the above considerations? Focussing on the Yukawa interaction we can first complexify $y$ 
\be
iyf \l e^{\frac{i\phi}{2f}}-e^{\frac{-i\phi}{2f}}\r \bar \psi \psi\Longrightarrow i \l fy e^{\frac{i\phi}{2f}}-fy^*e^{\frac{-i\phi}{2f}}\r \bar \psi \psi
\ee
and then extend it to a spurion in momentum space according to
\be
fy e^{\frac{i\phi}{2f}}\bar \psi\psi \Longrightarrow \left  [f\hat y\right ](p_1,p_2,p_3)\,
\left [e^{\frac{i\phi}{2f}}\right ](p_1)\, \,\bar \psi(p_2)\psi(p_3)\,.
\ee
We then need to focus only on two  symmetries, the fermion higher spin symmetry and the shift  $\phi\to \phi + \alpha f$ under which respectively 
\bea
\left  [f\hat y\right ](p_1,p_2,p_3)&\longrightarrow & \left [f\hat y\right ] y(p_1,p_2,p_3) e^{-i(\theta_\psi(p_2)+\theta_\psi(p_3))}\\
\left  [f\hat y\right ](p_1,p_2,p_3)&\longrightarrow &\left  [f\hat y\right ] y(p_1,p_2,p_3) e^{-i\alpha/2}
\eea
Notice that we are disregarding the $\phi$ higher  spin symmetry, under which $e^{i\phi/2f}$ transforms in a very complicated way. The above two symmetries then are easily seen to jointly allow a renormalization of the potential of the form
\be
\delta V\propto (f y)^2 M^2 e^{\frac{i\phi}{f}}+{\mathrm{h.c.}}
\ee
but they offer no clue as  to what $M$ should be. In principle it could be the UV cut-off, represented in our model by  $m_\rho$. However  direct computation shows the role of $M$ is instead played by another physical and lower  mass scale, $\sim 4\pi f$, controlling the onset of multiparticle processes. The mathematical reason for this result is the  exponential UV decrease of the two point function
\be
\langle e^{\frac{i\phi(x)}{f}} e^{\frac{i\phi(0)}{f}}\rangle\propto e^{-\frac{1}{4\pi^2 f^2 x^2}}\,.
\ee
This behaviour, as far as we can tell, is just dictated by unitarity rather than by simple dimensional analysis or some other symmetry.


\section{Dressing a non-Abelian Yukawa interaction}

In this section we {\it very} briefly discuss one manner in which our results might be applied to a non-abelian theory.  Consider the spontaneous symmetry breaking of $SU(N_f)/SU(N_f-1)$ via a scalar in the fundamental representation $\vec \Pi = e^{i \pi^a T^a/f_\pi} \cdot \vec \Pi_0$.  If one wishes, some of the $\pi^a$ may be loosely interpreted as Higgs like particles.  We couple these pNGBs to a fermion in the fundamental representation $\vec \Psi$ and a singlet $\Psi^c$.  Finally we have the pNGB of a $U(1)$ symmetry $\phi$, which will act as regulator of the quadratic divergence.

The Yukawa coupling we consider has the form 
\bea \label{Eq: nonabelian Yukawa}
\delta \mathcal{L} = y \vec \Psi \cdot (\vec \Pi \, e^{i \phi/f_\phi} - \vec \Pi^\dagger \, e^{-i \phi/f_\phi} ) \Psi^c.
\eea
This Yukawa coupling explicitly breaks $SU(N_f)$ down to $SO(N_f)$.  Additionally, the manner in which $\phi$ appears is enforced by the $U(1)$ symmetry.  As this coupling respects $SO(N_f)$, the pions of $SO(N_f)/SO(N_f-1)$ will decouple from the story while the remaining pNGBs will obtain a Yukawa coupling with the fermions and subsequently a mass as well.

Eq.~(\ref{Eq: nonabelian Yukawa}) resembles eq.~(\ref{eq: IR EFT}) except with $y \,\vec \Pi$ appearing out front instead of the combination $y \, f$.  To avoid doing any new calculations, we will be taking the limit $f_\phi \ll \luv = m_\rho \ll f_\pi$.  In this limit, we can sum over the many loop diagrams involving $\phi$ that give the exponentiation while {\it not} summing over loops involving the pions $\pi^a$.  We can thus treat $y \,\vec \Pi$ as a background field and simply repeat the calculation done in previous sections.  At order $y^2$ and $\l \luv/f_\pi \r^0$, the above Yukawa coupling generates a mass term for the pNGBs of the form
\bea
\delta V \propto  \l 4 \pi f_\phi \r^2 \, \l \vec \Pi \cdot \vec \Pi \r \,  e^{2 i \phi/f_\phi} + h.c.
\eea
with the calculation proceeding exactly as before.  Due to the dressing of the Yukawa coupling and mass term by $\phi$, we see that the mass term for the pNGBs is regulated by $4 \pi f_\phi$ as opposed to $\luv$.  While not a true non-abelian implementation of what was seen in this paper, we hope that this example may provide inspiration for the reader.

\bibliographystyle{JHEP}
\bibliography{Refs}

\end{document}